\documentclass[aip,graphicx, reprint,11pt]{revtex4-1}

\usepackage{graphicx}
\usepackage{dcolumn}
\usepackage{bm}
\usepackage{amsmath}
\usepackage{subfigure}
\usepackage[utf8]{inputenc}
\usepackage[T1]{fontenc}
\usepackage{etoolbox}

\makeatother
\begin{document}


\title[Beam-Plasma Dynamics in Finite-Length, Collisionless Inhomogeneous Systems]{Beam-Plasma Dynamics in Finite-Length, Collisionless Inhomogeneous Systems}

\author{R. Mishra}
\address{ 
Department of Physics, University of Oslo, PO Box 1048 Blindern, NO-0316 Oslo, Norway
}%
\email{rinku.mishra@fys.uio.no}
\author{R. Moulick}
\address{Centre of Plasma Physics, Institute for Plasma Research, Nazirakhat, Sonapur 782402, Assam, India.}
\address{Homi Bhaba National Institute, Training School Complex, Anushaktinagar, Mumbai, 400094, India}
\author{S. Adhikari, S. Marholm}
\address{Department of Computational Materials Processing, Institute for Energy Technology, Instituttveien 8, Kjeller 2007, Norway}
\author{A. J. Eklund}
\address{Materials Physics Oslo, SINTEF Industry, Forskningsveien 1, NO-0373 Oslo, Norway}
\author{ W. J. Miloch}
\address{ 
Department of Physics, University of Oslo, PO Box 1048 Blindern, NO-0316 Oslo, Norway
}%

\begin{abstract}
This study investigates the streaming instability triggered by ion motion in a plasma system that is finite in length, collisionless, and inhomogeneous. Employing numerical simulations using Particle-In-Cell (PIC) techniques and kinetic equations, the study examines how inhomogeneity emerges from integrating a cold ion beam with a background plasma within a confined system. The findings suggest that steady ion flow can modify ion sound waves through acoustic reflections from system boundaries, leading to instability. Such phenomena are known to be a hydrodynamic effect. However, there are also signatures of the beam-driven ion sound instability where kinetic resonances play a pivotal role. The main objective is to understand the impact of a finite-length system on beam-plasma instability and to identify the wave modes supported in such configurations.
\end{abstract}

\maketitle

\section{\label{sec:level1}{\textbf{Introduction}}}

Streaming instability\cite{briggs1964electron} is one of the oldest and most elementary examples of collective instabilities in plasma physics. For example, astrophysical plasmas containing small populations of non-thermal particles are susceptible to strong beam-plasma instabilities, leading to redistribution of energy among non-thermal populations. Examples include but are not limited to, various astrophysical phenomena, such as active galactic nuclei (AGNs) driven beam–plasma instabilities \cite{birk2001hard,broderick2014implications,lyubarsky2001reconnection}, the solar wind \cite{ginzburg1958possible}, pulsar wind outflows \cite{kirk2003dissipation,weiler1978crab}, shock formation and the associated acceleration of cosmic rays \cite{kang2002acceleration,schlickeiser2013cosmic}. \\
Moving on to bounded systems, there is a large relevance of beam–plasma interactions in intense heavy ion beams for applications to ion-beam-driven high energy density physics (HEDP) and heavy ion fusion (HIF) \cite{logan2007recent,davidson2006us,kaganovich2007charge,davidson2009survey}. Even after several decades of research, the complex nature of the instability and a great deal of open questions continue to draw the attention of the plasma community. \\
The concept of collective beam-plasma interaction was first introduced by Langmuir \cite{langmuir1925scattering} in 1925. However, it took a couple of decades to finally realize the potential of this specific phenomenon. In 1948, Pierce \cite{pierce1948possible} and later in the same year, Haeff \cite{haeff1948space} showed the prospects of two-stream instability in signal amplification. In the following year, Bohm and Gross \cite{bohm1949theory1,bohm1949theory2} theorized the problem as an electrostatic phenomenon in the absence of any magnetic field with kinetic equations. In the past several decades, the problem has been explored by numerous scientists \cite{stix1962theory,stringer1964electrostatic,davidson1999kinetic,davidson2000effects}.\\
Most of the works highlighted above deal with electron beams in a plasma. Since the ion-beam-induced gas ionization is faster than that of electron-beam in the same setting, the dynamics are different when an intense ion beam is introduced in a charge-neutralizing background plasma \cite{humphries1981one,sudan1984neutralization,oliver1996evolution}.  Similar situations can be found in experiments where plasma beams are introduced inside chambers filled with neutral gas. The beam interacts with the background neutrals and produces a plasma that provides nearly complete charge neutralization of the beam \cite{welch2002simulations,rose1999numerical}. A complete description of such systems can be described by nonlinear Vlasov-Maxwell equations \cite{davidson2001physics,reiser1994theory}. However, if we consider the effect of inhomogeneity and intense self-field, it becomes extremely difficult to predict the beam equilibrium, stability, and transport properties. In this manuscript, we do not intend to deal with relativistic beams; hence self-field can be ignored. However, the presence of electrically conducting boundaries such as a typical plasma chamber wall causes inhomogeneity, and it has a significant impact on the nature of streaming instabilities. Such instabilities appear to be more absolute than convective, much like a Pierce diode \cite{pierce1944limiting,pierce1948possible}.

In the absence of any external or self-generated magnetic field, the most significant process among the collective processes associated with an ion beam and a charge-neutralizing background plasma is the electrostatic electron-ion two-stream instability \cite{startsev2007dynamic}. The source of free energy to drive the classical two-stream instability comes from the relative streaming of beam electrons through the charge-neutralizing background plasma. The nature of such beam-plasma interactions is found to be highly dependent on the system length. In infinite systems, for the simplest case where collisions are neglected, and the ions do not have any drift velocity, the growth of such waves occurs at a certain frequency of the order of electron plasma frequency\cite{marholm2022current}. For finite-length systems with small drift velocities for background ions, the wave appears to grow at frequencies of the order of the ion plasma frequency instead \cite{pierce1948possible}. The inclusion of collisions and increasing temperature certainly affects the growth and effectively bounds the amplification rate \cite{boyd1958excitation}. However, it is worth mentioning, that the present work does not consider any effect of collision.

In systems where ion beams stream through thermal electrons, in relatively cold plasmas, the growth of waves is found to be strongest near the electron plasma frequency. However, it can also be of the order of the ion plasma frequency, provided the unperturbed beam velocity is very small compared to the average thermal velocity of background plasma electrons \cite{rukhadze1962electromagnetic,kitsenko1962excitation}. Such non-equilibrium systems may also exhibit excitation of ion-sound waves when the relative velocity between the ion stream and background electrons exceeds the ion sound velocity ($\Delta v > c_s$) \cite{rapson2014effect,skiff2001ion}. 

There are limitless implications of instabilities due to streaming ion beams such as in electric propulsion systems \cite{koshkarov2015ion}, plasma diodes \cite{ender2000collective,schamel1993lagrangian}, double layers \cite{johnson1990double,keesee2005ion}, and even for the sheath region at plasma-material boundaries \cite{farouki1990boundary}, Depending on the triggering mechanism, it can either lead to hydrodynamic instabilities \cite{koshkarov2015ion, wesson1973ion} or kinetic instabilities \cite{davidson2004collective,startsev2007dynamic,rose2007two}. 

In our present work, we present a collective picture of instabilities due to ion beams streaming through thermal plasma in a bounded system. In complex systems such as ours, the most difficult part is to address the effect of inhomogeneity due to the presence of the boundary in the system. The ideal theory of beam-plasma instability does not stand well when we bring all the factors into the picture. Upon examining the effects of inhomogeneity, it is found that wave modes exhibit significant variations with increasing levels of inhomogeneity. There are several studies associated with beam-plasma instability that address inhomogeneous medium \cite{shalaby2020growth,ziebell2010nonlinear,magelssen1977nonrelativistic}. Apart from the classical papers by Takakura and Shibahashi \cite{takakura1976dynamics}, Magelssen and Smith \cite{magelssen1977nonrelativistic}, the recent works by Shalaby et al. \cite{shalaby2020growth,shalaby2018growth} are worth mentioning in this regard. However, these papers address the inhomogeneity assuming the electron/positron beams streaming within inhomogeneous media. The most relevant work with the closest proximity would be the work by Rapson et al. \cite{rapson2014effect}, where they studied the effect of boundaries on the ion-acoustic beam-plasma instability. The present paper stands out in two different aspects. First, we assume a cold ion beam streaming through thermal electrons and background ions. Secondly, the inhomogeneity in the system arises naturally due to the presence of conducting surfaces on both ends. \\
One of the important aspects of such studies is to understand the plasma waves excited in ionospheric simulators. The ionospheric simulators are plasma chambers equipped with ion sources of high Mach number flow in a neutralizing environment. These chambers can reproduce ionospheric plasma conditions to test electrical probes before they go out for space missions. The present paper aims to explain the instabilities found in such devices to improve the control environment for space probes.\\
The paper is organized as follows. In section II, dispersion theory for homogeneous and inhomogeneous medium are presented followed by numerical methods and simulation setup details in section III. Results and relevant discussions are provided in section IV. Finally, in section V the work is summarized and concluded.

\section{Analytical theories of wave dispersion in beam-plasma systems}
\subsection{{\label{sec:2a}}Dispersion theory for homogeneous medium}
For the ion beam-driven ion acoustic instability, the dispersion relation for the simplest one-dimensional case takes the following form \cite{fitzpatrick2014plasma,jones2012introduction,nakamura2004effects},
\begin{equation}
    \begin{aligned}
        \epsilon(k, \omega)=1+\frac{e^{2}}{\epsilon_{0} m_{e} k} \int_{u} \frac{\partial F_{0 e} / \partial u}{\omega-k u} du \\+\frac{e^{2}}{\epsilon_{0} m_{i} k} \int_{u} \frac{\partial F_{0 i} / \partial u}{\omega-k u}du \\ +\frac{e^{2}}{\epsilon_{0} m_{i} k} \int_{u} \frac{\partial F_{0ib} / \partial u}{\omega-k u}du =0
    \end{aligned}
\label{eq:1}
\end{equation}

where, $m_e$, and $m_i$ are the mass of electron and ion respectively. $~\epsilon_0,~e$ represents the permittivity of free space and elementary charge. $k$ stands for the wave number and $\omega$ is the wave frequency. $F_{0 e}$, $F_{0 i}$, and $F_{0 ib}$ are the phase space distributions of background thermal electrons, thermal ions, and cold beam ions respectively.

For a wave with a phase velocity $\omega/k$ much less than electron thermal velocity but larger than ion thermal velocity, we can assume $\omega \ll ku$ for background electrons and $\omega \gg ku$ for background ions. Hence, the second term in the right-hand side of \ref{eq:1} can be approximated as,

\begin{equation}
\frac{e^{2}}{\epsilon_{0} m_{e} k} \int_{u} \frac{\partial F_{0 e} / \partial u}{\omega-k u} du = \frac{\omega_{pe}^2}{k^2}\frac{m_e}{T_e} = \frac{1}{k^2\lambda_D^2}
\label{eq:2}
\end{equation}
considering that the background electrons are Maxwellian with a temperature $T_e$, $$F_{0e}(u)=\frac{n_e}{\left(2 \pi T_{e} / m_{e}\right)^{1 / 2}} \exp \left(-\frac{m_{e} u^{2}}{2 T_{e}}\right),$$
the electron plasma frequency: $\omega_{pe}=\sqrt{\frac{n_e e^2}{\epsilon_0 m_e}}$, the Debye length: $\lambda_{D}=\sqrt{\frac{\epsilon_0 k_b T_e}{n_e e^2}}$, $n_e$ is electron density, and $k_b$ represents Boltzmann constant.

Similarly, the third term can be approximated as,
\begin{equation}
    \frac{e^{2}}{\epsilon_{0} m_{i} k} \int_{u} \frac{\partial F_{0 i} / \partial u}{\omega-k u} du = -\frac{\omega_{pib}^2}{\omega^2}
\label{eq:3}   
\end{equation}
where, ion plasma frequency: $\omega_{pib}=\sqrt{\frac{n_{ib} e^2}{\epsilon_0 m_i}}$

For the ion beam contribution, we can not consider $\omega \gg ku$, as the beam velocity ($v_b$) is comparable to the wave velocity. Hence, the fourth term can be approximated as,

\begin{equation}
    \frac{e^{2}}{\epsilon_{0} m_{i} k} \int_{u} \frac{\partial F_{0 ib} / \partial u}{\omega-k u} du = -\frac{\omega_{pib}^2}{{(\omega-kv_{b})}^2}
\label{eq:4}   
\end{equation}
Using, eq. \ref{eq:2} to eq. \ref{eq:4} in eq. \ref{eq:1} we get,

\begin{equation}
    \epsilon(k, \omega)= 1 + \frac{1}{k^2\lambda_D^2} -\frac{\omega_{pi}^2}{\omega^2} -\frac{\omega_{pib}^2}{{(\omega-kv_{b})}^2} = 0
\label{eq:5}  
\end{equation}
The first three terms represent the bulk plasma oscillations and the last term comes due to the presence of the beam. The frequencies of the excited wave modes are of the order of ion time scale ($\omega_{pi}$).
The normalized eq. \ref{eq:5} takes the following form,
\begin{equation}
\epsilon(\Tilde{k}, \Tilde{\omega})= 1 + \frac{1}{\Tilde{k}^2} -\frac{1}{\Tilde{\omega}^2} -\frac{\alpha}{{(\Tilde{\omega}-\Tilde{k}\Tilde{v_{b}})}^2} = 0
\label{eq:6}  
\end{equation}
$$\Tilde{k} = k \lambda_D, ~~ \Tilde{\omega} = \frac{\omega}{\omega_{pi}}, ~~\Tilde{v_{b}}=v_{b}/C_s$$
where, $\Tilde{k}$, $\Tilde{\omega}$, $\Tilde{v_{b}}$ and $\alpha = n_{ib}/n_i$ are the normalized wave number, normalized plasma frequency, normalized beam velocity and the ion beam density. $C_s = \sqrt{T_e/m_i}$ is the ion acoustic speed.

We aim to use the dispersion theory to draw conclusions for complex cases from PIC simulations where we expect to have kinetic effects. Such comparison will help in isolating the effects from kinetic properties of beam plasma systems and to assess the finite-length effects on such systems.

\subsection{{\label{sec:2b}}Dispersion theory for beam-plasma in inhomogeneous medium}

The formalism for estimating the growth of a beam-plasma instability in an inhomogeneous medium is naturally complicated as there are many factors in play. In this section, we adopt the formalism developed by Shalaby et al. \cite{shalaby2020growth} to estimate the growth in such a system. In the present case, the first-order (linearized) Vlasov-Poisson equations can be re-written as an eigenvalue problem as:

\begin{equation}
    \begin{aligned}
        \frac{e^{2}}{\epsilon_{0} m_{e} k} \int_{u}\int_{k'} \frac{\partial F_{0 e}(k',u) / \partial u}{\omega-k u} E_{1}\left(k-k^{\prime}, \omega\right) dk'du\\
        +\bigg[1+\frac{e^{2}}{\epsilon_{0} m_{i} k} \int_{u} \frac{\partial F_{0 i} / \partial u}{\omega-k u} du \\
        +\frac{e^{2}}{\epsilon_{0} m_{i} k} \int_{u} \frac{\partial F_{0 ib} / \partial u}{\omega-k u} du \bigg] E_{1}(k, \omega)=0
    \end{aligned}
\label{eq:7}
\end{equation}

where $E_{1}$ is the first-order perturbation in the electric field. The important difference between eq. \ref{eq:7} and eq. \ref{eq:1} is the inhomogeneity introduced in the background electron term. $F_{0e}$ is the equilibrium phase space distribution of thermal electron of the following form,
\begin{equation}
    F_{0 e}(x,u) = n_{e}(x)F_{0 e}(u)
\label{eq:8}
\end{equation}
Assuming the plasma inhomogeneity to be quadratic, the background density for electrons can be expressed as,
\begin{equation}
    n_{e}(x) = n_{e0}(1+\xi x^{2}); ~\xi \geq 0
\label{eq:9}
\end{equation}
where $\xi$ is the inhomogeneity factor and has a dimension of inverse length squared and 
Using eq. \ref{eq:9} in eq. \ref{eq:8} and taking the Fourier transform, the equation becomes,
\begin{equation}
F_{0 e}\left(k^{\prime}, u\right)=n_{e0}\left[\delta\left(k^{\prime}\right)-\xi \delta^{\prime \prime}\left(k^{\prime}\right)\right] F_{0 e}(u)
\label{eq:10}
\end{equation}
Using eq. \ref{eq:10} in the first term of eq. \ref{eq:7} and integrating over $k'$ gives,
\begin{equation}
    \begin{aligned}
        \bigg[\frac{e^{2}}{\epsilon_{0} m_{e} k} \int_{u} \frac{\partial F_{0 e} / \partial u}{\omega-k u} du\bigg] (1 - \xi \partial_{k}^{2}) E_{1}(k, \omega)
    \end{aligned}
\label{eq:11}
\end{equation}
Splitting eq. \ref{eq:11} and simplifying we obtain,
\begin{equation}
    \begin{aligned}
        \frac{e^{2}}{\epsilon_{0} m_{e} k}  \int_{u} \frac{\partial F_{0 e} / \partial u}{\omega-k u} du E_{1}(k, \omega) \\
        -\xi \frac{e^{2}}{\epsilon_{0} m_{e} k}  \int_{u} \frac{\partial F_{0 e} / \partial u}{\omega-k u} du \partial_{k}^{2} E_{1}(k, \omega)
    \end{aligned}
\label{eq:12}
\end{equation}

Now, using eq. \ref{eq:2}, the integrals in eq. \ref{eq:12} can be rewritten as,
\begin{equation}
\begin{aligned}
& \left(\frac{1}{k^2\lambda_D^2}\right) E_{1}(k, \omega)
 -\xi \left(\frac{1}{k^2\lambda_D^2}\right) \partial_{k}^{2} E_{1}(k, \omega)
\end{aligned}
\label{eq:13}
\end{equation}
Using eq. \ref{eq:3} to eq. \ref{eq:4} and eq. \ref{eq:13}, we can rewrite eq. \ref{eq:7} as,
\begin{equation}
\begin{aligned}
&\epsilon (k,\omega) = \\& \left(\frac{1}{k^2\lambda_D^2}\right) E_{1}(k, \omega) 
-\xi \left(\frac{1}{k^2\lambda_D^2}\right) \partial_{k}^{2} E_{1}(k, \omega) \\&
+\left[1 -\frac{\omega_{pi}^2}{\omega^2} -\frac{\omega_{pib}^2}{{(\omega-kv_{b})}^2}\right]E_{1}(k, \omega) = 0
\end{aligned}
\label{eq:14}
\end{equation}



Now, eq. \ref{eq:14} can be rewritten as,
\begin{equation}
\begin{aligned}
&-\frac{ \xi}{ k^2 \lambda_D^2}\partial_{k}^{2} E_{1}
+\left[ 1 -\frac{\omega_{pi}^2}{\omega^2} -\frac{\omega_{pib}^2}{{(\omega-kv_{b})}^2} \right]E_{1} \\& +\frac{1}{ k^2 \lambda_D^2} E_{1}=0
\end{aligned}
\label{eq:15}
\end{equation}
Similar to eq. \ref{eq:6}, if we write eq. \ref{eq:15} using normalized expressions for $\omega$, $k$, and $\xi$,
\begin{equation}
\frac{ -\Tilde{\xi}}{\Tilde{k}^2} \partial_{\Tilde{k}}^{2} E_{1}+  \left[ 1 -\frac{1}{\Tilde{\omega}^2} -\frac{\alpha}{{(\Tilde{\omega}-\Tilde{k}\Tilde{v_{b}})}^2} \right]E_{1} + \frac{ 1}{\Tilde{k}^2}E_{1}=0
\label{eq:16}
\end{equation}
where, $\Tilde{\xi} = \xi \lambda_{D}^2$, and $E_1$ is a function of $\Tilde{k},\Tilde{\omega}$.

\begin{equation}
    -\frac{A}{\tilde{k}^2}\frac{\partial^2E_1}{\partial\tilde{k}^2} + \left(B + \frac{1}{\tilde{k}^2}\right)E_1 = 0
\end{equation}
where, $A = \Tilde{\xi}$, and $B = \left[ 1 -\frac{1}{\Tilde{\omega}^2} -\frac{\alpha}{{(\Tilde{\omega}-\Tilde{k}\Tilde{v_{b}})}^2} \right]$

In the limit of small perturbation to the potential by the beam, using the solution for Weber differential equations \cite{zwillinger1998handbook},  the solution for first order perturbation in the E-field can be written as,
\begin{equation}
E_1(\Tilde{k})=C_{1} D_{\nu}(y)+C_{2} D_{-\nu-1}(iy)
\label{eq:17}
\end{equation}
where, $\nu = {-\frac{1}{2}-\frac{1}{2 \sqrt{A}\sqrt{B}}} $,~ $y = \frac{\sqrt{2} \sqrt[4]{B} \Tilde{k}}{\sqrt[4]{A}}$, 
$D_{\nu}(y)$ is the parabolic cylinder function \cite{whittaker2020course,weisstein2002parabolic}. $C_1$ and $C_2$ are the constants of integration. For real values of $\nu$ and $y$ the function $D_{\nu}(y)$ will be real.

We consider $\nu$ to be a non-negative integer it can be denoted as $n$. Then $D_{\nu}(y)$ (ignoring the complex solution part) can be rewritten as ,
\begin{equation}
    D_n(y) = 2^{-n/2}e^{-y^2/4}H_n\left( \frac{y}{\sqrt{2}}\right), ~ n \in Z^+
\label{eq:18}
\end{equation}
where $H_n$ is a Hermite polynomial

Now, using \ref{eq:18}, the general expression for \ref{eq:17} can be written as,
\begin{equation}
E_1(\Tilde{k},\Tilde{\omega}) = C_12^{-n/2} e^{-y^2/4}H_n\left( \frac{y}{\sqrt{2}}\right)
\label{eq:19}
\end{equation}
Substituting eq.\ref{eq:19} in eq.\ref{eq:16} and eliminating common factors on both sides we get,
\begin{equation}
-(n+\frac{1}{2})\frac{\Tilde{\xi}}{\Tilde{k^2}} + \frac{1}{2}\sqrt{\Tilde{\xi}}\sqrt{B}+ B + \frac{ 1}{\Tilde{k}^2} = 0
\label{eq:20}
\end{equation}
From eq.\ref{eq:20}, we can see as the inhomogeneity ($\Tilde{\xi}$) approaches zero, the dispersion takes the shape of eq.\ref{eq:6} representing homogeneous periodic beam plasma system.

\section{\label{sec:modeling}Numerical method and simulation setup}

\begin{figure}[ht]
\centering
\includegraphics[width=8.0cm]{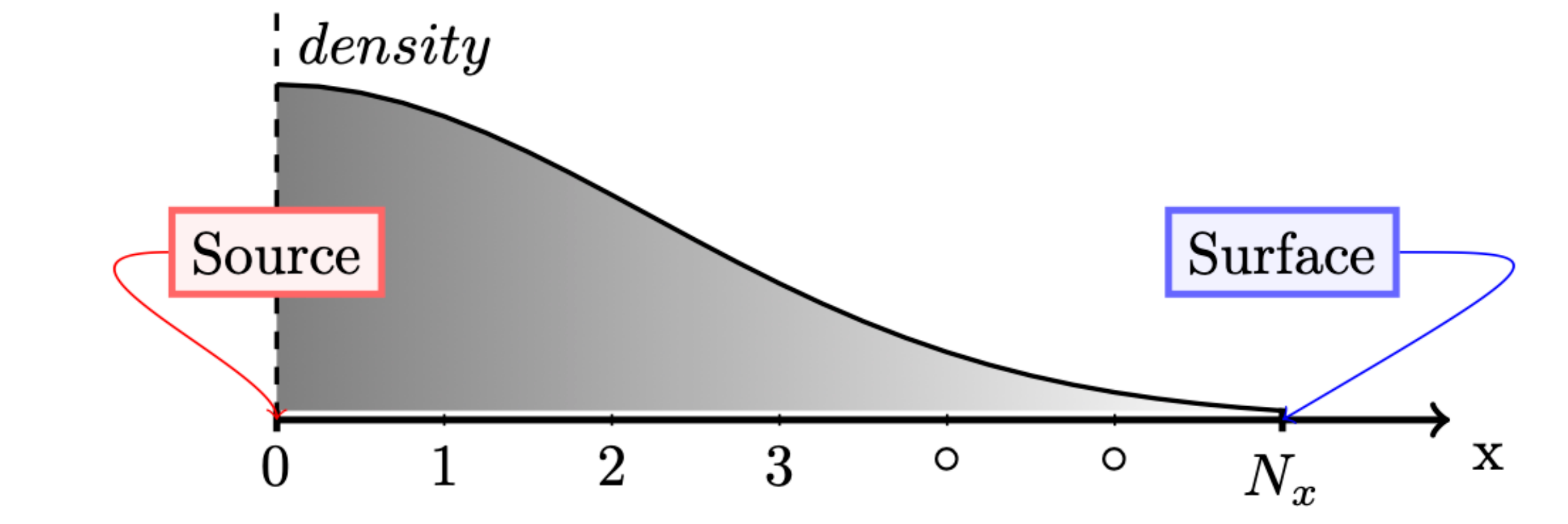}
\caption{\label{fig:1} Schematic of the simulation model}
\end{figure}

Kinetic simulations are performed with a 2D3V particle-in-cell (PIC) code, XOOPIC (X11-based Object-Oriented Particle-In-Cell) \cite{verboncoeur1995object}.The model uses two-dimensional Cartesian geometry to represent a plasma system of variable lengths (see figure\ref{fig:1}). The system is assumed periodic in the $y$-direction, reducing the problem to 1D. A plasma source is introduced in the system from the left-hand side of the domain. 
The source is modeled such that ions can have different drift velocities. We have considered ions to be cold ($0.01 ~eV$), background ions with no drift and ion beam with drift velocities, ranging from $0.1 ~eV$ to $7.5 ~eV$ which in normalized units, is equivalent to $\sim 0.489~C_s$ to  $\sim 4.242~C_s$. Electrons are introduced with only the thermal component at $0.5 ~eV$. The specific values for system parameters have been adopted from the ionospheric environment simulator system at the University of Oslo, Norway. The plasma is considered unmagnetized, and the self-generated magnetic field is ignored, assuming small current values. The plasma density has been taken as $\approx 10^{13}~m^{-3}$, and hydrogen has been considered to be the ion species. The ratio of ion beam to ion density is $\alpha = 0.1$. Field quantities are dumped periodically to measure the temporal growth rate of instability as well as the dispersion of the electrostatic waves. In all of the case studies, the cell size ($\Delta x$) is taken as the electron Debye length ($\lambda_D$) of the respective system. The time step is considered sufficiently small to resolve electron plasma oscillations. In order to study finite-length effects, the system length has been investigated for different lengths $512 ~\lambda_D$, $1024 ~\lambda_D$, and $2048 ~\lambda_D$. The baseline numerical parameters have been used from the table I.

\begin{table}[h]
\centering
\begin{tabular}{  c  c  }
    Parameters & Value\\
    \hline
    Plasma density ($N_{e,i}$) &  $ \sim 10^{13}~m^{-3}$  \\
    Electron temperature ($T_e$) &  0.5~eV \\ 
    Ion temperature ($T_i$)  &  0.01~eV   \\ 
    Ion beam temperature ($T_{ib}$)  &  0.01~eV   \\
    Debye length ($\lambda_d$) & $0.001$ m\\
    Number of  cells ($N_x$, $N_y$) & 1024, 4  \\ 
    Spatial Grid size ($\delta x$, $\delta y$) & $0.001$ m, $0.001$ m \\
    System length ($L_x$, $L_y$) & 1.7 m, 0.006 m\\
    Time step ($\Delta t$)  & 0.02$\omega_{pe}^{-1}$\\
    No. of particles      & 1048576
\end{tabular}
\label{table:1}
\caption{Numerical simulation parameters.}
\end{table}

Although the code accepts unnormalized input, the outputs have been normalized using relevant quantities to emphasize the physical scaling. Ion velocities are normalized with ion sound speed ($C_s$), whereas electron velocities are normalized with the electron thermal velocity ($v_{th}$). Normalized ion and electron velocities are expressed as $\Tilde{v_i}$, and $\Tilde{v_e}$ respectively. Time is normalized with the inverse of ion plasma frequency ($\omega_{pi}$) and denoted as $\tau$. Lengths are normalized with electron Debye length ($\lambda_{D}$) and expressed as $\Tilde{x}$. Lastly, kinetic energies ($\varepsilon_{i,e}$) are normalized by the thermal energy ($\varepsilon_{i,e}^{th}$) of the respective species.

\section{\label{sec:results}Results and discussions}
The main objective of this investigation is to establish the role of ion flow in the development of streaming instabilities in inhomogeneous bounded systems. It is expected that bounded systems act very differently as compared to periodic systems and theoretical justifications for such cases are hard to establish. In order to understand the parameter range unaffected by boundary effects, we initially developed an analytical framework applicable to both homogeneous and inhomogeneous media. The results for the infinite/periodic system are not included, as the scope of this work was confined to a bounded system.

Additionally, investigating the relevance of the inhomogeneous dispersion theory in the present context requires various factors to be taken into account. One major simplification that has been introduced was neglecting pressure impacts by setting the temperature to zero, meaning the thermal velocity ($v_{th}$) is much less than the beam velocity ($v_{b}$). Later in the paper, we highlight how this simplification helps in pinpointing the wave modes.

\subsection{\label{sec:dispersion_theory}Results from kinetic dispersion theory}

\begin{figure*}
    \subfigure[]{\includegraphics[width=0.45\linewidth]{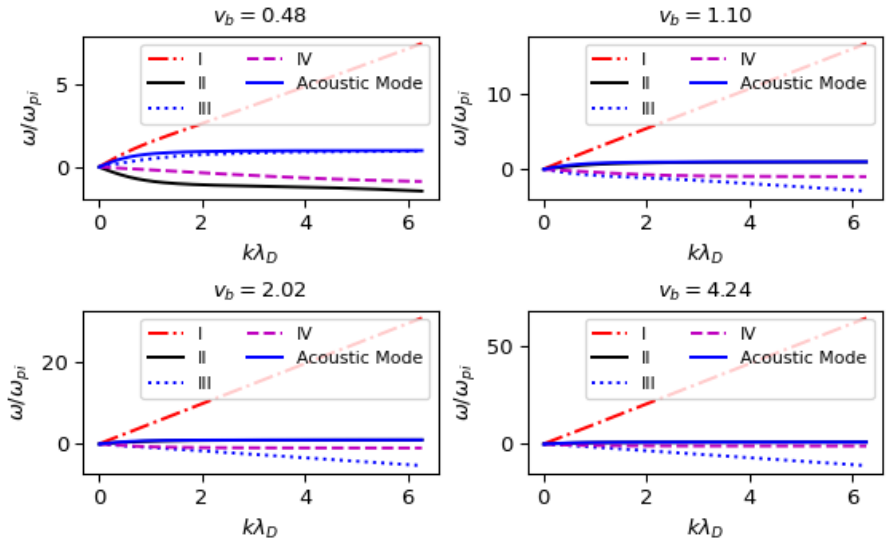}}
    \hfill
    \subfigure[]{\includegraphics[width=0.45\linewidth]{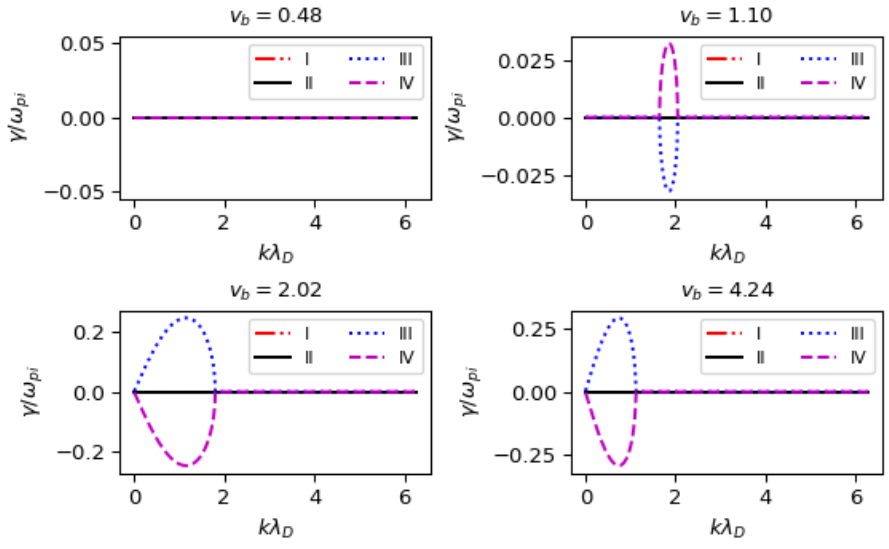}}
    \vspace{0.5cm}
    \subfigure[]{\includegraphics[width=0.45\linewidth]{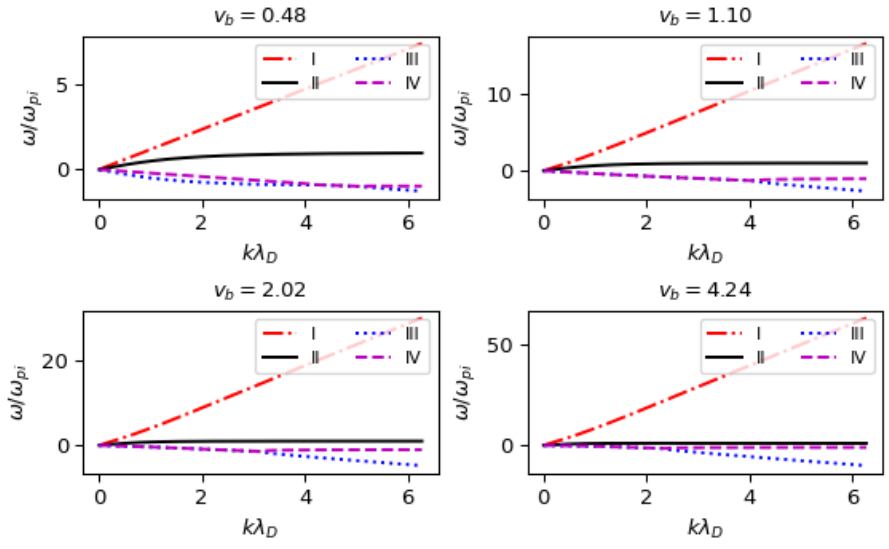}}
    \hfill
    \subfigure[]{\includegraphics[width=0.45\linewidth]{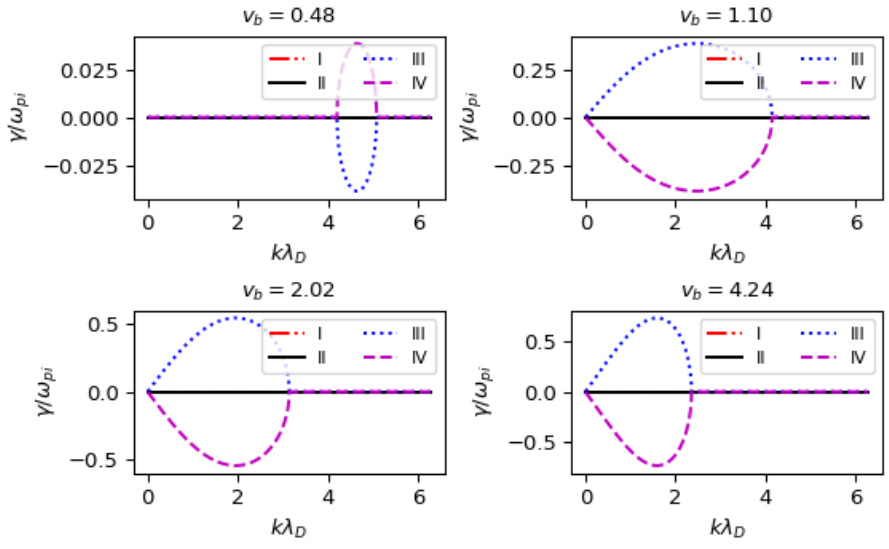}}

\small{\caption{\label{fig:2} Analytic solutions for the inhomogeneous bounded system from \ref{eq:20} with (a) - (b) $\Tilde{\xi} = 0.08$, (c) - (d) $\Tilde{\xi} = 1.5 $. Four different wave modes represented by I, II, III, IV. (b) and (d) shows the growth and damping of the wave modes.  }}

\end{figure*}

Figure.\ref{fig:2} shows the solutions for different wave modes with subsonic, sonic, and supersonic beam velocities in the presence of small and large inhomogeneity in the system. The four roots of the dispersion relation I, II, III, and IV has been obtained by solving the equation \ref{eq:20} for $n= 0$ (shown in figure \ref{fig:2} (a and c)). Ion acoustic mode (blue solid line) are from the theory of ion acoustic wave shown here for reference. The numerical parameters we have used here are from the experimental set of the ionospheric simulator present at the University of Oslo. One of the significant outcomes from this is the presence of a ion-acoustic branch (figure \ref{fig:2} (a and c)). Interaction of the ion beam with the background plasma system shows the presence of unstable modes. 

Inhomogeneity plays a big role in the growth of the wave modes. For strong inhomogeneity, the wave modes gets more damped for higher beam velocity. The wave modes appear to be coupled at low $\Tilde{k}$ and get separated for the higher values of the wave vector. When there is a small inhomogeneity present, the analysis shows that the wave modes have zero imaginary parts for the subsonic beam velocity(figure \ref{fig:2}(b)). Whereas the presence of imaginary components is observed in the roots II and IV, indicating instability within the system, for the ion beam exhibiting the sonic and supersonic velocities. Furthermore, the wave growth remains consistent even with increasing the velocities of the ion beam than supersonic values. Additionally, a correlation has been identified between the enhancement of inhomogeneity values and the emergence of unstable modes in the subsonic beam velocities (figure \ref{fig:2} c and d). Growth and damping of the wave mode is also seemed to be have reasonable effect on the ion beam velocity. \\
As the inhomogeneity gets stronger, with $\Tilde{k}$ the wave undergoes a reflection as is expected in a bounded system. In the remainder of this paper, we will see how this could explain the results from kinetic simulations. 

\subsection{Results from kinetic simulations}
To explore the effect of the ion beam on the development of instability within the system, our model employs a two-component plasma, consisting of the ion, ion beam and electrons. We will consider the effects of the finite-length system on the evolution of beam-plasma instability. Since the ion acoustic waves are compressional in nature, the presence of a boundary will lead to a significant modification in the plasma properties. To differentiate the effects induced by the boundaries, along with the analytic dispersion theory of inhomogeneous medium (section \ref{sec:2b}), we have performed kinetic simulations of the same set-up introducing boundaries and beam as a source at one of the boundaries. 

\begin{figure}
\centering
\includegraphics{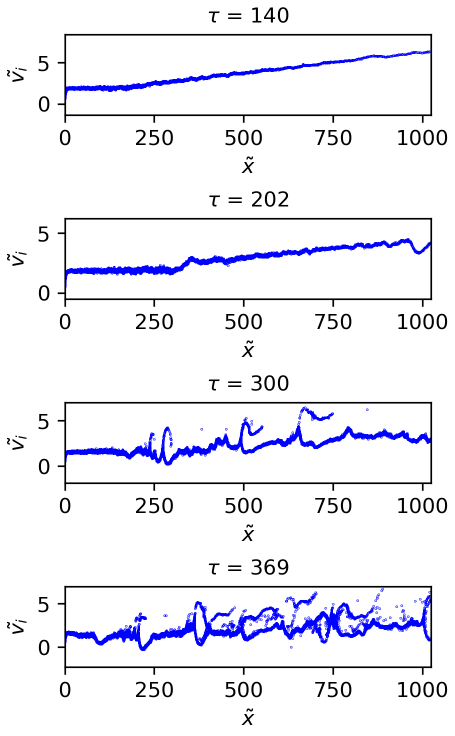}
\caption{\label{fig:3} Phase space of ions at different stages of simulation for the bounded system with $\Tilde{v_b} = 0.489$ and $\Tilde{x} = 1024$. Times are mentioned inside each panel.}
\end{figure}

\begin{figure}
\centering
\includegraphics{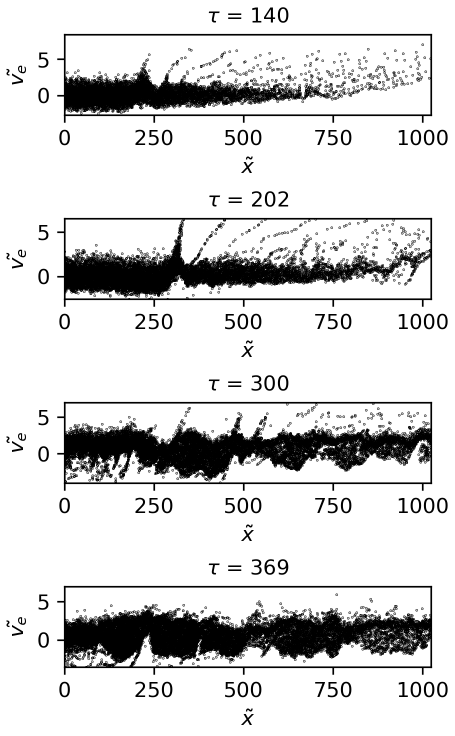}
\caption{\label{fig:4} Phase space of electrons at different stages of simulation for the bounded system with $\Tilde{v_b} = 0.489$ and $\Tilde{x} = 1024$. Times are mentioned inside each panel.}
\end{figure}

\begin{figure}
\centering
\includegraphics[width=0.5\textwidth]{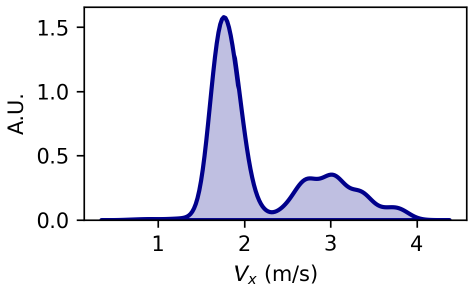}
\caption{\label{fig:5} Distribution of ions for bounded systems with $\Tilde{v_b} = 0.489$ and $\Tilde{x} = 1024$, $\tau =500$. The distribution shown here represents the distribution after observing the instability in the phase space.}
\end{figure}

\begin{figure}
\centering
\includegraphics[width=0.5\textwidth]{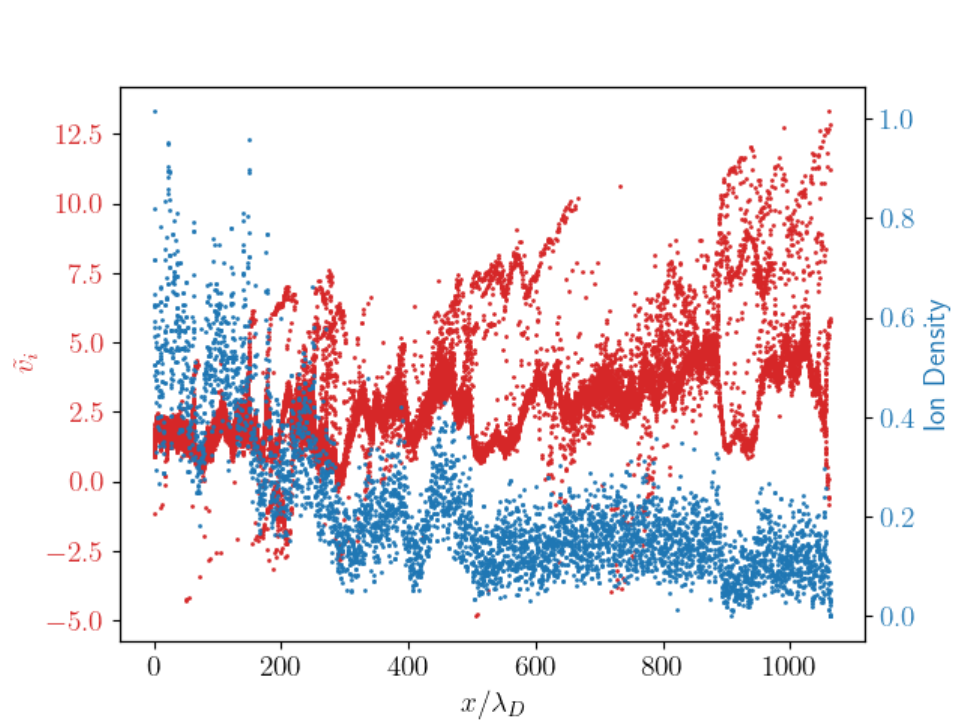}
\caption{\label{fig:6} Ion density(blue) and phase velocity(red) of ions for $\Tilde{v_b} = 0.489$ and $\Tilde{x} = 1024$. This data has been considered for $\tau=800$. Density is normalised with ions equilibrium density and velocity with ion acoustic speed.}
\end{figure}

In the simulation model particles have been injected system from the left boundary of the domain. The right boundary is assumed to be a perfectly absorbing conductor, i.e. as soon as any particle hits the right boundary it is removed from the domain. For such systems, we run simulations for a enough longer time to reach a sufficiently developed state to study the growth of the irregularities. For all of our cases, we set the limit at $\tau = 500$. In our simulation setup, observations were made until the system attained a steady state. It is seen that particle interactions, both among themselves and with the boundary, result in energy loss, leading to the formation of a subset of particles exhibiting varied energy levels. Consequently, within the analytical framework, we focused on that steady-state scenarios in which an ion beam propagates through a system that includes ions and electrons as the background plasma.

\begin{figure}
\centering
\includegraphics{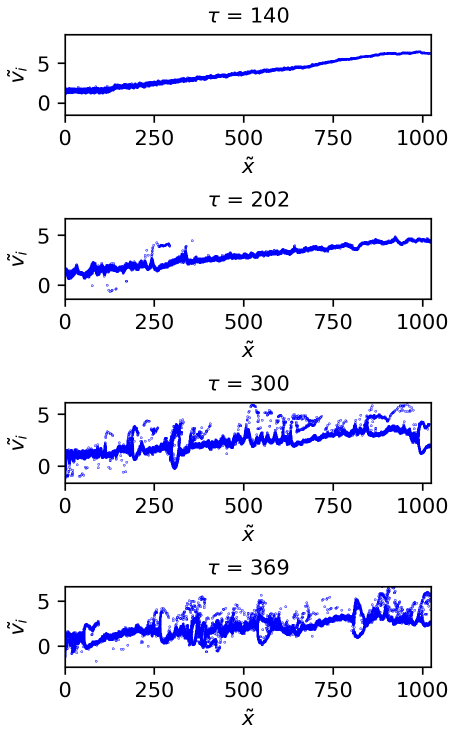}
\caption{\label{fig:7} Phase space of ions at different stages of simulation for the bounded system with $\Tilde{v_b} = 1.0$ and $\Tilde{x} = 1024$. Times are mentioned inside each panel.}
\end{figure}

\begin{figure}
\centering
\includegraphics{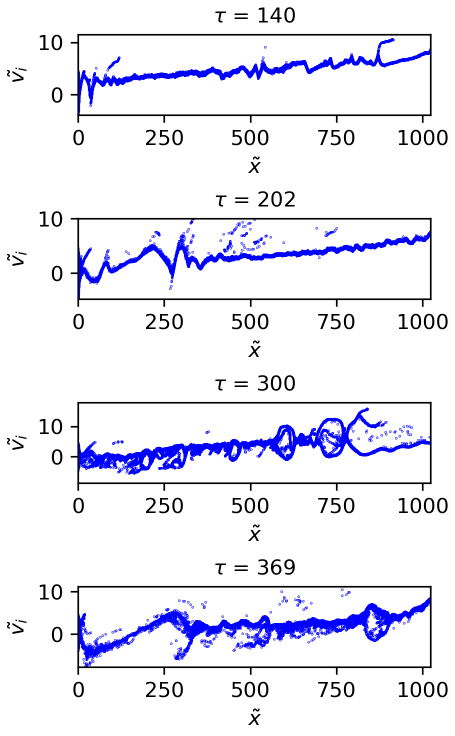}
\caption{\label{fig:8} Phase space of ions at different stages of simulation for the bounded system with $\Tilde{v_b} = 4.242$ and $\Tilde{x} = 1024$. Times are mentioned inside each panel.}
\end{figure}

The figure \ref{fig:3} and \ref{fig:4} show the phase space structures of individual species over time for system length $\Tilde{x} = 1024$. A significant observation has been made from the phase spaces of individual species. The particles launched from the left reach the boundary, and the acoustic branch present in the system undergoes reflection resulting in the decelerating particles with different energy modes. Therefore this reflection created the situation where particles with accelerating and decelerating nature are present. The kinetic interaction between these particles makes the system unstable \cite{koshkarov2015ion}. The reflected particle with decelerating nature forms the populations of trapped ions, leading to the formation of an ion hole (as illustrated in figure \ref{fig:3}). To understand the dynamics of particles following reflection, we analyzed the distribution of the ion beam (figure \ref{fig:5}), and found that that a segment of the particle population exhibits ion velocities exceeding the initially provided input velocity.  Furthermore, analysis of the ion phase space and density we noticed, the region encompassing the ion hole is characterized by a thin population of ion particles. This can be clearly seen in the figure \ref{fig:6} for the region near to the boundary (around $\tilde{x} \sim 950$). From figure \ref{fig:5}, it can be concluded that these thin populations within the ion distribution are playing a pivotal role in the creation of an ion hole. In the figure \ref{fig:7} and \ref{fig:8} we have plotted the phase space of ions for $\tilde{v_b} = 1.0$ and $\tilde{v_b} = 4.242$. In the figure we can see that for both the beam velocity interaction between the reflected and the initial particles started at a very early stage like for $\tilde{v_b} = 1.0$ $\tau=202$ and for $\tilde{v_b} = 4.242$ $\tau=140$. The another interesting observation is that even after increasing the beam velocity higher than $\tilde{v_b} = 4.242$ there are not much significant changes in the phase space. This observation is very much similar to our analytical results.

Next we investigated the energy of the particle with different varying system length to see the time of emergence of instability. In figure \ref{fig:7}, we can see a significant energy growth in the average electron energy at $\tau \sim 100 $ and around $\tau \sim 200$. Here, one can make an important observation in terms of the beam acceleration. As soon as the beam enters the system, it starts to accelerate toward the boundary and loses its energy to the background plasma particles.

One of the important mechanisms behind ion sound instability in bounded systems is charge separation. As we increase the system length, the charge separation becomes less prominent, resulting in a decreasing instability growth rate \cite{koshkarov2015ion}. For long systems where $\lambda_{D} \ll L$, the medium is considered weakly dispersive i.e. $k\lambda_{D} \ll 1$. For a fixed-length system, the instability growth rate is a function of the ion beam velocity ($\Tilde{v_b}$).

\begin{figure}
\centering
\includegraphics{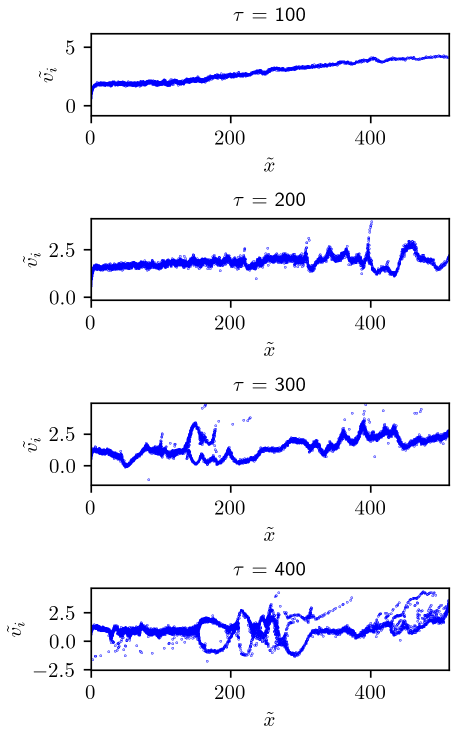}
\caption{\label{fig:9} Phase space of ions at different stages of simulation for a bounded system with $\Tilde{v_b} = 0.489$ and $\Tilde{x} = 512$. Times are mentioned inside each panel.}
\end{figure}

\begin{figure}
\centering
\includegraphics{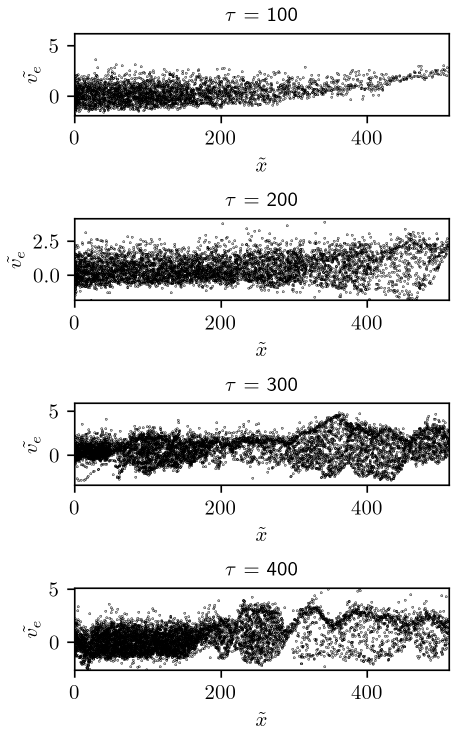}
\caption{\label{fig:10} Phase space of electrons at different stages of simulation for a bounded system with $\Tilde{v_b} = 0.489$ and $\Tilde{x} = 512$. Times are mentioned inside each panel.}
\end{figure}

Fig. \ref{fig:9} and \ref{fig:10} show the phase space structure of ions and electrons over time for the system length $\Tilde{x} = 512$. In comparison with \ref{fig:3} and \ref{fig:4}, the dependency of instability growth on the boundary is visible: in the former case the phase space holes are more prominent at an earlier stage ($\tau \sim 202$).

\begin{figure*}
    \subfigure[]{\includegraphics[width=0.45\linewidth]{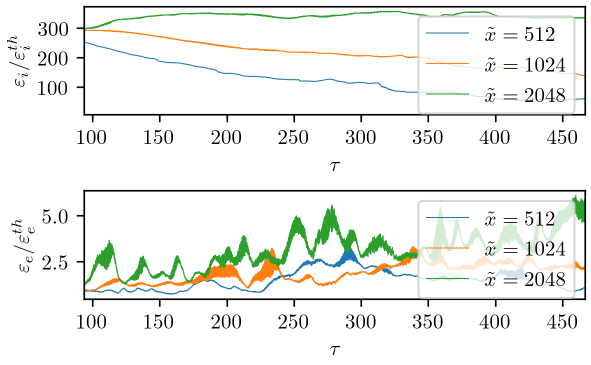}}
    \hfill
    \subfigure[]{\includegraphics[width=0.45\linewidth]{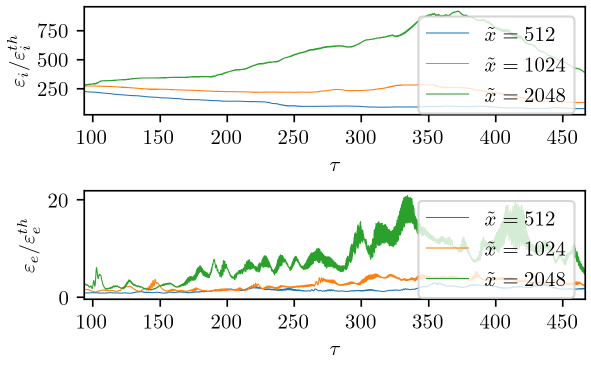}}
    \vspace{0.5cm}
    \subfigure[]{\includegraphics[width=0.45\linewidth]{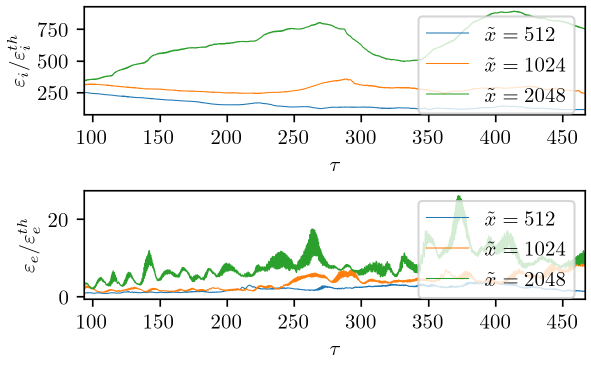}}
    \hfill
    \subfigure[]{\includegraphics[width=0.45\linewidth]{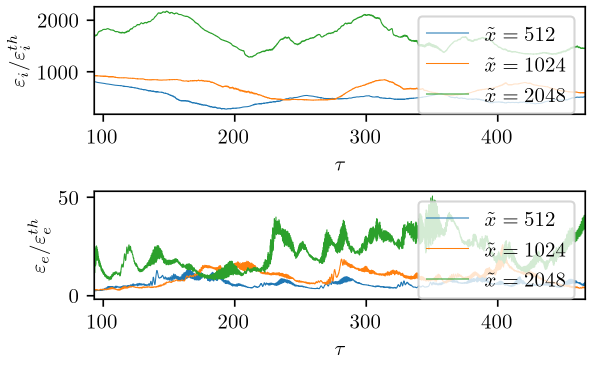}}

\small{\caption{\label{fig:11} Evolution of the average kinetic energy of ions (top) and electrons (bottom) for bounded systems with (a) $\Tilde{v_b} = 0.489$, (b) $\Tilde{v_b} = 1.095$, (c)$\Tilde{v_b} = 1.549$, and (d) $\Tilde{v_b} = 4.242$. The average kinetic energy of individual species is normalized by the thermal energy of the respective species.}}

\end{figure*}

Figure \ref{fig:11} gives us an overview of the evolution of the average kinetic energy of individual species for different system lengths ($\Tilde{x} $) and beam velocities ($v_b$). For the subsonic case (see figure\ref{fig:11} (a)) with lengths $\Tilde{x} = 512$, and $\Tilde{x} = 1024$, the average kinetic energies for electrons and ions seem to follow the same trend. At around $\tau \sim 100$, as the instability starts to develop in the system ($\Tilde{x} = 512$), the ion energy falls rapidly and transfers energy to electrons. One important observation for the sonic case is that it takes a longer time to develop instability for the larger system length.
For sonic and supersonic cases (see figure \ref{fig:11}b and \ref{fig:11}c), due to the higher beam velocity, the system starts to support the wave growth faster and the energy becomes oscillatory. For the near highly supersonic case (see figure \ref{fig:9} (d)), we did not notice any difference in the trend. 
From figure\ref{fig:9} it is clear that the growth is stronger for systems with longer lengths as compared to shorter ones similar to the case reported by Koshkarov et al. \cite{koshkarov2015ion}. 

The stronger growth for the longer system can be explained in terms of average kinetic energies for particles (see figure \ref{fig:11}). For shorter systems, the instability is triggered faster, and in due process, the ions start to lose their energy to the electrons. Therefore, the wave modes start to disperse faster as compared to the larger systems. Due to the presence of the conducting boundary, the particles are removed as soon as they hit the boundary leading to a self-consistent formation of the sheath. The presence of a sheath influences the particle flow to the boundary. The shorter system gets affected strongly in contrast to longer systems. In figures \ref{fig:3} and \ref{fig:9}, we can see that for the same $\Tilde{v_b}$, the ion velocities are limited to $\sim 3$ for a shorter system, whereas ions have higher velocities ($\sim 5$) in the larger system.

\section{\label{sec:summary}Summary and Conclusions}
The present work addresses the effects of finite-length systems on beam plasma instability for a cold ion-beam streaming through a background plasmas. The study reveals a significant influence of boundary conditions on the excitation of wave modes within plasmas. The ion acoustic waves has been observed across all beam velocities. In the current working model, using specific numerical parameters,  neither growth nor damping of the wave was observed for subsonic beam velocities when the inhomogeneity was small. However, with an increase in inhomogeneity, unstable modes became apparent even at subsonic beam velocities. Additionaly, we also observed the increase of wave frequency amplitude of one of the wave modes for increasing beam velocity.

The phase spaces for ions in bounded systems appear very interesting as it indicates the presence of ion hole. One of the important reasons for such nature is thought to be the acoustic reflection at the boundary which can destabilize the sound waves. The interaction of the accelerated an decelerated particles forms the ion hole in which it is assumed that the particles with decelerating nature gets trapped in the hole. These trapped particles have higher energy as compared to the initial particles. For increasing the beam velocity to sonic and supersonic region, the interactions between the particles occurs at the early stage. From the phase space investigation we found the similarity between the analytical and simulation results. Increasing the beam velocity beyond supersonic levels does not markedly impact the system. From this observation, we can conclude that our chosen numerical parameters yield results that agree with both simulation and kinetic outcomes.

Inhomogeneous plasmas are non-trivial to address. In the present work, alongside theory, we have been able to simulate the effects of boundaries on beam plasma instability. The implications of the present work will greatly help in understanding the wave modes in ionospheric plasma simulator devices (e.g. plasma devices at the University of Oslo, Norway, and at ESTEC in the Netherlands \cite{bekkeng2010design}) or any devices with ion sources (e.g. ion thrusters, hollow cathodes, field effect emitters, plasma contactors, etc.) streaming through a thermal neutralizing environment.

One of the critical outcomes of this study is the ability to quantify the beam plasma instability. Extensive parameter scans for such systems allow us to configure the modes and control the plasma for several applications. The insights are applicable to experimental systems involving ion beams injected in background plasmas in bounded systems, and provide a framework to understand observations of beam-plasma instability. The dynamics are complex and require detailed numerical simulations, possibly with parameter sweeps, to quantitatively match and explain experimental results.

\begin{acknowledgments}
This study is a part of the 4DSpace Strategic Research Initiative at the University of Oslo.
This work received funding from the European Research Council (ERC) under the European Union’s Horizon 2020 research and innovation program (Grant Agreement No. 866357, POLAR-4DSpace).\\ The work has also been supported by the Research Council of Norway (RCN), grant numbers: 275653, 275655, and 325074. The authors would also like to thank Bjørn Lybekk for his helpful advice on various technical issues for data management in simulations.\\
The simulations were performed on resources provided by Sigma2 - the National Infrastructure for High Performance Computing and Data Storage in Norway, project number NN9987K.
\end{acknowledgments}

\section*{Data Availability Statement}
The data that support the findings of this study are available from the corresponding author upon reasonable request.

\section*{References}
\bibliographystyle{unsrt}
\bibliography{bibliography}

\end{document}